# All-passive, transformable optical mapping (ATOM) near-eye display

WEI CUI,[1,2] AND LIANG GAO[1,2,*]

[1]Department of Electrical and Computer Engineering, University of Illinois at Urbana-Champaign, 306 N. Wright St., Urbana, IL 61801, USA
[2]Beckman Institute for Advanced Science and Technology, University of Illinois at Urbana-Champaign, 405 N. Mathews Ave., Urbana, IL 61801, USA
*Corresponding author: gaol@illinois.edu

**We present an all-passive, transformable optical mapping (ATOM) near-eye display based on the "human-centric design" principle. By employing a diffractive optical element, a distorted grating, the ATOM display can project different portions of a 2D display screen to various depths, rendering a real 3D image with correct focus cues. Thanks to its all-passive optical mapping architecture, the ATOM display features a reduced form factor and low power consumption. Moreover, the system can readily switch between a real-3D and a high-resolution 2D display mode, providing task-tailored viewing experience for a variety of VR/AR applications.**

http://dx.doi.org/XX.XXXX/OL.XX.XXXXXX

The emergence of virtual reality (VR)/augmented reality (AR) technologies has revolutionized the way that people access the digital information. Despite tremendous advancement, currently, very few VR/AR devices are crafted to comply with the "human-centric design" principle [1], which puts human perception, wearability, and usability in the center of hardware design and serves as the blueprint for the near-eye displays' future [2]. To meet this gold standard, a near-eye display must coherently integrate displays, sensors, and processors, while allowing for human-computer interaction in a compact enclosure. Among these four pillar requirements, the display plays a central role in creating a three-dimensional (3D) perception that mimics real-world objects.

Conventional near-eye 3D displays are primarily based on computer stereoscopy [3], presenting two images with parallax in front of the viewer's eyes. Stimulated by binocular disparity cues, the viewer's brain then creates an impression of the three-dimensional structure of the portrayed scene. However, the stereoscopic displays suffer from a major drawback, the vergence-accommodation conflict [4], which reduces the viewer's ability to fuse the binocular stimuli while causing discomfort and fatigue. Because the images are displayed on one surface, the focus cues specify the depth of the display screen (*i.e.*, accommodation distance) rather than the depths of the depicted scenes (*i.e.*, vergence distance). This is opposite to the viewer's perception in the real world where these two distances are always the same.

To alleviate this problem, we have recently developed an optical mapping near-eye (OMNI) display which can project a volumetric image directly onto the viewer's retina, ushering the wearable displays into a new era [5]. Notwithstanding the "real 3D" display capability, the OMNI display relies on an active optical component, liquid-crystal-on-silicon spatial light modulator (LCOS-SLM), to execute its core function, unfavorably increasing power consumption and the device's form factor and therefore jeopardizing the device's wearability.

To enable a "real-3D" display that conforms to the human-centric design principle, herein we present an all-passive, transformable optical mapping (ATOM) near-eye display method. Based on a conceptual thread similar to the OMNI display, the ATOM display simultaneously maps different portions of a two-dimensional (2D) display screen to various depths while forcing their centers aligned. However, rather than using the LCOS-SLM, the ATOM display employs a passive diffractive optical element—a distorted grating—to achieve 2D-to-3D mapping, reducing the power consumption and the device's form factor. Moreover, to improve the device's usability, we build the system on a transformable architecture which allows a simple switch between the real-3D and high-resolution 2D display modes, providing task-tailored viewing experience.

We illustrate the operating principle of the ATOM display in Fig. 1. In the real-3D display mode, we divide the input screen into multiple sub-panels, each displaying a depth image. These images are then relayed by a $4f$ system with a distorted grating at the Fourier plane. Acting as an off-axis Fresnel lens, the distorted grating adds both the linear and quadratic phase factors to the diffracted waves, directing the associated sub-panel images to a variety of depths while shifting their centers towards the optical axis. Seeing through the eyepiece, the viewer will perceive these sub-panel images appearing at different virtual depths. Also, by rendering the contents using a depth-blending algorithm [6], we can provide continuous focus cues across a wide depth range.

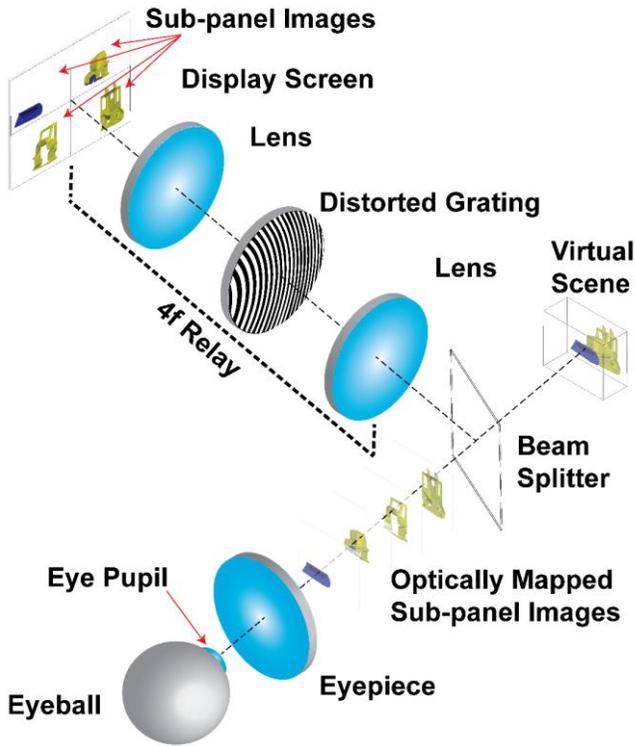

Fig. 1. Operating principle.

Due to the division of the display screen, given $N$ depth planes, the resolution of each depth image is therefore $1/N$ of the display screen's native resolution, leading to a reduced field of view (FOV). To accommodate applications where a large FOV is mostly desired, we can transform the ATOM display into a high-resolution 2D display simply by removing the distorted grating from the optical path and displaying a single plane image at the screen's full resolution. This switching mechanism thus grants users a freedom to adapt the ATOM display for a specific task.

We implemented the ATOM display in the reflection mode. The optical setup is shown in Fig. 2. At the input end, we used a green-laser-illuminated digital light projector (DLP4500, 912×1140 pixels, Texas Instruments) as the display screen. After passing through a 50:50 beam splitter, the emitted light is collimated by an infinity-corrected objective lens (focal length, 100mm; 2X M Plan APO, Edmund Optics). In the real-3D display mode, we place a reflective distorted grating at the back aperture of the objective lens to modulate the phase of the incident light. While in the high-resolution 2D display mode, we replace the distorted grating with a mirror. The reflected light is collected by the same objective lens, reflected at the beam splitter, and forms intermediate depth images (real-3D display mode) or a full-resolution 2D image (high-resolution 2D display mode) in front of an eyepiece (focal length, 8 mm; EFL Mounted RKE Precision Eyepiece, Edmund Optics). The resultant system parameters for the high-resolution 2D and real-3D display modes are shown in Table 1.

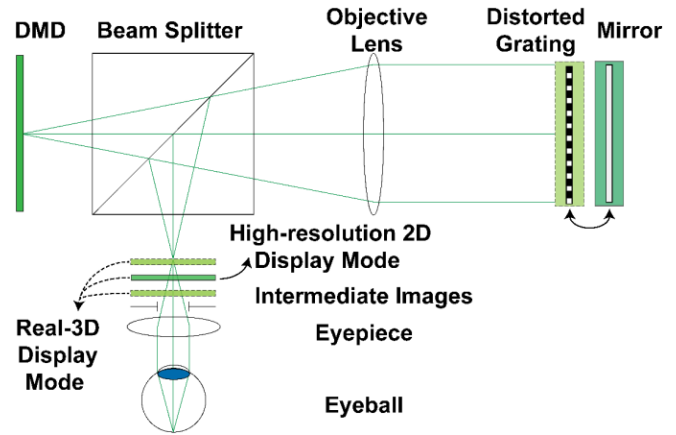

Fig. 2. Optical schematic. DMD, digital mirror device.

Table 1. System parameters of an ATOM display.

|  | Resolution (pixels) | FOV (degrees) |
|---|---|---|
| High-resolution 2D display mode | 900×900 | 63 |
| Real-3D display mode | 300×300 | 23 |

As an enabling component, the distorted grating functions as a multiplexed off-axis Fresnel lens in the ATOM display. Although distorted gratings have been long used in microscopy, wavefront sensing, and optical data storage [7-10], we deploy it for the first time in a display system. We elaborate the effect of a distorted grating on an optical system in Fig. 3(a). Given a single object, the distorted grating introduces varied levels of defocus to the wavefront associated with different diffraction orders. When combined with a lens, the distorted grating modifies its focal length and laterally shifts the image for non-zero diffraction orders. Similarly, given multiple objects located at the same plane but different lateral positions, the distorted grating can simultaneously project their different diffraction-order images onto various depths while maintaining their centers aligned (Fig. 3(b)).

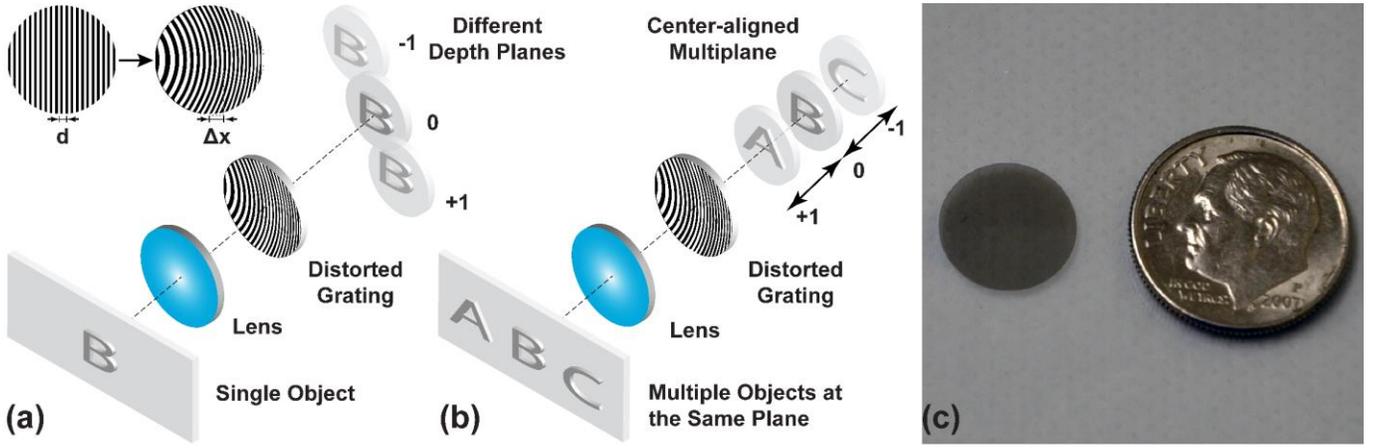

Fig. 3. Image formation in a distorted-grating-based optical system. (a) Diffraction of a single object through a distorted grating. (b) Diffraction of multiple in-plane objects through a distorted grating. Only the on-axis diffracted images are illustrated. (c) Photograph of a distorted grating. A US quarter is placed at the right for size reference.

The unique diffraction property above originates from the spatially-varied shift in the grating period, $\Delta_x(x,y)$ (Fig. 3(a)). The correspondent local phase shift for diffraction order $m$ can be written as:

$$\varphi_m(x,y) = \frac{2\pi m \Delta_x(x,y)}{d} + \frac{2\pi m x}{d}, \quad (1)$$

where $d$ is the period of an undistorted grating. At the right side of Eq. 1, the first and second term depict the contributions from the distorted and undistorted grating period, respectively. If the first distorted term has a quadratic form,

$$\Delta_x(x,y) = \frac{W_{20} d}{\lambda R^2}(x^2 + y^2), \quad (2)$$

where $R$ is the grating radius, and $W_{20}$ is the defocus coefficient, and $\lambda$ is wavelength, the correspondent phase change $\varphi_m^Q$ would be:

$$\varphi_m^Q(x,y) = m \frac{2\pi W_{20}}{\lambda R^2}(x^2 + y^2). \quad (3)$$

We can consider this phase change is contributed by a lens with an equivalent focal length,

$$f_m = \frac{R^2}{2 m W_{20}}. \quad (4)$$

The sign of diffraction order $m$ thus determines the optical power of the distorted grating.

On the other hand, the second undistorted term in Eq. 1 introduces a linear phase shift to the wavefront in the form:

$$\varphi_m^L(x,y) = \frac{2\pi \sin\theta}{\lambda} x, \quad (5)$$

where $\theta$ is the diffraction angle, and it can be calculated from the grating equation:

$$d \sin\theta = m \lambda. \quad (6)$$

Under the small-angle approximation, we correlate the diffraction angle $\theta$ with the lateral shift $l_{x_m}$ of a sub-panel image in the ATOM display as:

$$\sin\theta = \frac{l_{x_m}}{f_{OBJ}}, \quad (7)$$

where $f_{OBJ}$ is the focal length of the objective lens in Fig. 2.

Finally, combining Eq. 1-7 gives:

$$\varphi_m(x,y) = \frac{\pi(x^2+y^2)}{\lambda f_m} + \frac{2\pi}{\lambda}\sin\left(\frac{l_{x_m}}{f_{OBJ}}\right)x. \quad (8)$$

Notably, the phase pattern in Eq. 8 is inherently associated with diffracted depth images. By contrast, in the OMINI display [5], to calculate the required phase pattern displayed at LCOS-SLM, we must perform optimization for each depth image, which is computationally extensive and may lead to an ill-posed problem when the number of depth planes increases.

In our prototype, we used only the $+1$, $0$, and $-1$ diffraction orders and projected their associated images to 0, 2, 4 diopters, respectively. The correspondent focal lengths were computed and shown in Table 2. We calculated the structural parameters of the distorted grating (Table 3) and fabricated it as a reflective mask using direct laser writing on a soda lime base with high reflective chrome coating (Fig. 3(c)).

Table 2. Calculated $f_m$ of an ATOM display.

| Diffraction order | +1 | 0 | -1 |
|---|---|---|---|
| Dioptric depth (diopter) | 0 | 2 | 4 |
| $f_m$ (m) | 81.4 | Inf. | -81.4 |

Table 3. Structural parameters of the distorted grating.

| $W_{20}$ (nm) | $d$ (μm) | $R$ (mm) |
|---|---|---|
| 185.8 | 43.2 | 5.5 |

To demonstrate the high-resolution 2D display, we captured a representative image at the intermediate image plane using a Sony Alpha 7S II digital camera (Fig. 4(a)). To evaluate the real-3D display performance, we performed a simple depth mapping experiment. At the input end, we displayed three letters "A", "B", "C" on the three sub-panels of the display screen respectively (Fig. 4(b)) and captured the remapped images at three nominal depth planes (0D, 2D, and 4D). To compensate for the intensity variation between 0 and ±1 diffraction-order images, we applied a neutral density filter to the central sub-panel image to dim its brightness. The remapped letter images at three designated depths are shown in Fig. 4(c)-(e), respectively. As expected, the letters appear sharp at their designated depths while blurred elsewhere.

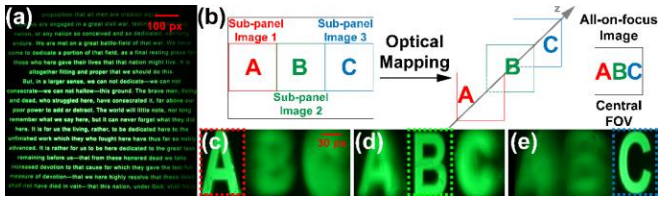

Fig. 4. Evaluation of display performance. (a) Image captured in the high-resolution 2D display mode. (b) Optical mapping in the real-3D display mode. (c-e) Images captured at three depths. px, pixels.

To assess the focus cues provided by the ATOM display, we measured the modulation transfer function (MTF) at different accommodation distances. We directly placed the camera at the nominal working distance of the eyepiece and varied its axial position to mimic the accommodation of an eye. Two identical sub-panel images (slanted edge) were displayed at the input screen and projected to 0D and 4D depth plane with their centers aligned. The rendered target depth is at the dioptric midpoint 2D position [11]. The experimentally-measured accommodation is 2.05D, closely matching with the target value (Fig. 5).

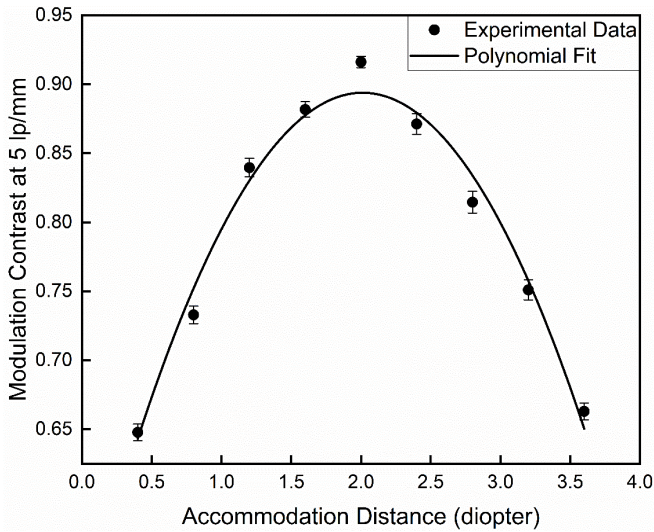

Fig. 5. Assessment of focus cues.

Next, we test the system's stability during the mechanical switch between two display modes. To characterize the tolerance of the distorted grating to the misalignment, we varied the distorted grating's position both laterally and axially and measured the correspondent display performance. Again, we chose MTF as the metric and used the dual-plane characterization method above. The results imply that the MTF decreases as the grating's position shift increases (Fig. 6). Here the position shift is calculated with respect to the grating's nominal position. Given a threshold ΔMTF =0.1, the system can tolerate a shift of 2mm along both lateral and axial axes. This relatively loose tolerance favors the low-cost production of the device.

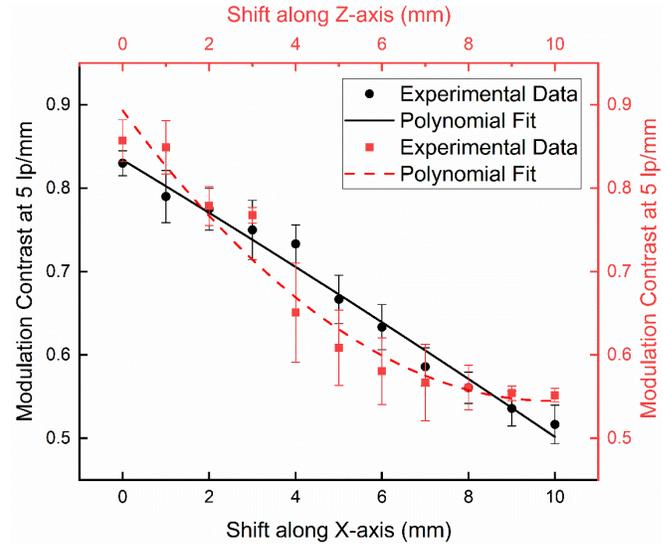

Fig. 6. Sensitivity of modulation contrast to the relative shift of the distorted grating.

Finally, we demonstrated the system's capability in displaying a complex 3D scene. To create a 3D scene with continuous focus cues, we employed a linear depth-weighted blending algorithm to render the contents of sub-panel images [12]. In brief, we set the image intensity at each depth plane proportional to the dioptric distance of the point from that plane to the viewer along a line of sight. Meanwhile, we maintained the sum of the image intensities a constant at all depth planes. To achieve uniform image brightness across the entire depth range, we applied a tent filter to the linear depth blending, where the light intensity for each depth plane reaches a maximum at its nominal position and minimum elsewhere.

Based on the algorithm above, we generated the sub-panel images for three nominal depth planes (0D, 2D, and 4D) for a 3D titled fence image and displayed them at the input screen. Then we placed a camera in front of the eyepiece, adjusted its focal depth to mimic the accommodation of the eye, and captured the images at a series of depths (Video 1). The representative depth-fused images at near-end (4D) and far-end (0D) are shown in Fig. 7(a) and (b), respectively, closely matching the ground-truth depth map (Fig. 7(c)).

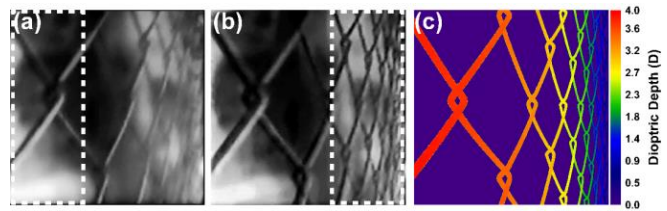

Fig. 7. ATOM display of a complex three-dimensional scene. Representative depth images captured at (a) near-end (4D) and (b) far-end (0D). (c) Ground-truth depth map.

In summary, based on the human-centric design principle, we developed a compact ATOM near-eye display which can provide correct focus cues that best mimic the natural response of human eyes. By projecting different sub-panel images of a 2D display screen to various depths using a distorted grating, we created a real 3D image over a wide depth range. The employment of all-passive optical components reduces the system dimension and power consumption, thereby improving the system's wearability. Moreover, the ATOM display can easily switch between a high-resolution 2D and real-3D display modes, providing task-tailored viewing experience and expanding its usability.

Although not demonstrated, we can enable more depth planes by using a distorted grating with periodic structures along two dimensions [13]. Using such a 2D diffractive element, we can perform lateral optical mapping along both $x$ and $y$ axes, leading to a more efficient utilization of screen pixels. In the ideal case, the total number of remapped pixels is equal to that of the original display screen. For example, given an input screen of $N \times N$ pixels, an ATOM display with a 2D distorted grating can project a total of nine depth images, each with $N/3 \times N/3$ pixels and associated with a unique diffraction order $(L_x, L_y)$, where $L_x, L_y = 0, \pm 1$.

In the current ATOM display prototype, we decrease the light intensity associated with 0 diffraction order to compensate for the difference in diffraction efficiency, however, at the expense of reduced light throughput. To fully utilize the dynamic range of the display screen, rather than using a binary-amplitude distorted grating, we can employ a sinusoidal-phase distorted grating [14] and build the system in the transmission mode. Such a diffractive phase element allows an approximately uniform energy distribution among $\pm 1$ and 0 orders, and it can be holographically fabricated by creating an interference pattern on a photoresist.

**Funding.** This work was supported in part by discretionary funds from UIUC and a research grant from Futurewei Technologies, Inc. The authors would also like to thank Ricoh Innovations for gifting the DMD modules.

## Full References